\documentclass[prl,aps,showpacs,nofootinbib,superscriptaddress,tightenlines,twocolumn]{revtex4}

\usepackage{epsfig}
\usepackage[latin1]{inputenc}
\usepackage{float,amsmath}
\usepackage{graphicx}

\newcommand{\beq}{\begin{equation}}
\newcommand{\eeq}{\end{equation}}
\newcommand{\bqa}{\begin{eqnarray}}
\newcommand{\eqa}{\end{eqnarray}}

\def\lsim{\mathrel{\rlap{\lower4pt\hbox{$\sim$}}
    \raise1pt\hbox{$<$}}}                
\def\gsim{\mathrel{\rlap{\lower4pt\hbox{$\sim$}}
    \raise1pt\hbox{$>$}}}                

\begin{document}

\title{Measuring Quark-Gluon-Plasma Thermalization Time with Dileptons}

\author{Mauricio Martinez}
\affiliation{Helmholtz Research School \\
  Johann Wolfgang Goethe - Universit\"at Frankfurt \\
  Ruth-Moufang-Stra\ss{}e~1,
  D-60438 Frankfurt am Main, Germany }
\author{Michael Strickland}
\affiliation{Institut f\"ur Theoretische Physik and Frankfurt Institute for Advanced Studies\\
  Johann Wolfgang Goethe - Universit\"at Frankfurt \\
  Max-von-Laue-Stra\ss{}e~1,
  D-60438 Frankfurt am Main, Germany }

\begin{abstract}

We calculate the medium dilepton yield from a quark-gluon plasma which 
has a time-dependent momentum-space anisotropy. A phenomenological 
model for the hard momentum scale, $p_{\rm hard}(\tau)$, and plasma 
anisotropy parameter, $\xi(\tau)$, is constructed which interpolates 
between longitudinal free streaming at early times ($\tau \ll 
\tau_{\rm iso}$) and ideal hydrodynamic at late times ($\tau \gg 
\tau_{\rm iso}$).  We show that high-energy dilepton production is 
sensitive to the plasma isotropization time, $\tau_{\rm iso}$, and can 
therefore be used to experimentally determine the time of onset for 
hydrodynamic expansion of a quark-gluon plasma and the magnitude of 
expected early-time momentum-space anisotropies.

\end{abstract} 
 \maketitle \newpage


An important question facing experimentalists and theorists working on 
heavy-ion experiments ongoing at the Relativistic Heavy Ion Collider 
(RHIC) and planned at the Large Hadron Collider (LHC) is at what time 
is it justified to assume that the matter created can be described 
using simple hydrodynamics.  At RHIC energies it has been found that 
for $p_T \lsim 2$ GeV the elliptic flow of the matter created is 
described well by models which assume ideal hydrodynamic behavior 
starting at very early times $\tau \lsim 1$ fm/c 
\cite{Teaney:2000cw,Huovinen:2001cy,Hirano:2002ds,Tannenbaum:2006ch}. 
Since then further refinements including viscous corrections have 
become available and indications are that the viscosity is constrained 
to be small \cite{Drescher:2007cd,Romatschke:2007mq,Song:2007fn} and 
results remain consistent with early thermalization of the quark-gluon 
plasma (QGP). However, given the complexity of solving three-dimensional 
viscous hydrodynamic equations coupled to a late-time 
hadronic cascade \cite{Nonaka:2006yn,Bass:1998ca} it would be nice to 
have an independent way to determine the time at which a QGP begins to 
undergo hydrodynamic expansion.  In this letter we propose to use 
high-energy dilepton yields as a function of both pair invariant mass and 
transverse momentum to experimentally determine QGP isotropization 
time.

The use of ideal hydrodynamics to describe matter requires at a 
minimum that the matter be isotropic in momentum space 
\cite{Arnold:2004ti}.  In practical applications it is also necessary 
to impose an additional constraint (equation-of-state, conformal 
invariance, etc.) in order to close the resulting system of 
hydrodynamic equations.  When an equation-of-state is applied it is 
implicitly assumed that the system is isotropic and thermal so that 
$\tau_{\rm iso} = \tau_{\rm therm}$. For simplicity we will also 
identify these two time scales. Estimates from perturbative QCD for 
the thermalization time of a QGP range from $2-3$ fm/c 
\cite{Baier:2000sb,Xu:2004mz,Strickland:2007fm}. Recently it has been 
shown that plasma isotropization is accelerated by unstable plasma 
modes induced by the rapid longitudinal expansion of the QGP fireball 
\cite{Mrowczynski:2000ed,Strickland:2007fm}; however, it is still not 
known by precisely how much.

In most phenomenological treatments of the QGP it is assumed that the 
plasma thermalizes rapidly with $\tau_{\rm iso} = \tau_{\rm therm}$ on 
the order of the parton formation time.  However, given the rapid 
longitudinal expansion of the matter this seems like a rather strong 
assumption and one would like to know the effect of momentum-space 
anisotropies on experimental observables. 

Absent a precise dynamical picture of the first few fm/c of the QGP's 
lifetime we propose a simple phenomenological model for the time-dependence 
of the plasma momentum-space anisotropy, $\xi = \frac{1}{2} \langle 
p_T^2 \rangle /\langle p_L^2 \rangle - 1$, and hard momentum scale, 
$p_{\rm hard}$. We then use this model to explore the effect of early-time 
plasma momentum-space anisotropies on high-energy dilepton 
production.  To accomplish this we introduce two parameters: (1) 
$\tau_{\rm iso}$ which is the proper time at which the system begins 
behaving hydrodynamically and (2) $\gamma$ which sets the sharpness of 
the transition to hydrodynamic behavior. For times greater than the 
parton formation time, $\tau_0$, but short compared to $\tau_{\rm 
iso}$ we will assume that the system is longitudinally free streaming 
and for times long compared to $\tau_{\rm iso}$ that it is expanding 
hydrodynamically.  To estimate the parton formation time we use the 
nuclear saturation scale, $\tau_0 \sim Q_s^{-1}$ 
\cite{Venugopalan:2007vb}. For RHIC energies $Q_s \simeq 1.5$ GeV and 
for LHC energies $Q_s \simeq 2$ GeV implying that $\tau_0 \simeq 0.2$ 
fm/c and $\tau_0 \simeq 0.1$ fm/c, respectively.

For our final results we present the invariant mass and transverse 
momentum dependence of medium dilepton production.  We compare our 
results with estimates of other relevant sources and background 
processes and demonstrate that increasing the QGP isotropization time 
from $\tau_{\rm iso} =\tau_0$ to $\tau_{\rm iso} = 2$ fm/c 
significantly increases medium high-energy dilepton production.  We 
present detailed calculations of this enhancement and show that at LHC 
energy it leads to an experimentally measurable effect on dilepton 
production between 3 GeV $< M, \, p_T <$ 8 GeV.

\begin{figure*}[t]
\includegraphics[width=17.7cm]{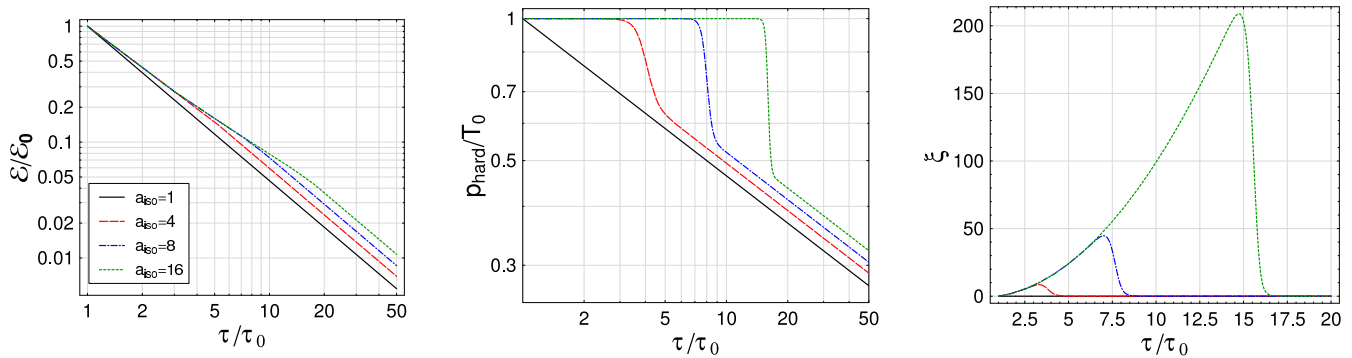}
\vspace{-2mm}
\caption{Model energy density (left), hard momentum scale (middle),
and anisotropy parameter (right) for four
different isotropization times $\tau_{\rm iso} \in \{0.1,0.4,0.8,1.6\}$ fm/c 
assuming $\tau_0 =$ 0.1 fm/c. 
The transition width is taken to be $\gamma = 2$.}
\label{fig:modelPlot}
\end{figure*}


\section{Dilepton Rate}
\label{sec:hardpart}

The leading order medium dilepton production rate comes from the 
annihilation process $q{\bar q} \rightarrow l^+ l^-$.  Medium dilepton 
production is computed by folding this rate together with the expected 
local quark and anti-quark phase space distribution functions 
\cite{Kajantie:1986dh,Kapusta:1992uy}. In this work we allow these 
phase space distribution functions to be anisotropic in momentum-space 
and then model the time-dependence of the anisotropy.  Assuming 
azimuthal symmetry about the beam ($z$) axis the time-dependence can 
be parameterized as
\beq
f_{\{ q ,{\bar q}\}}({\bf k},\tau) = 
   f_{ \{ q ,{\bar q}\} , \rm iso} 
     \left(\sqrt{{\bf k}^2 + \xi(\tau) \, k_z^2} \, , \;
       p_{\rm hard}(\tau) \right) ,
\label{eq:fansio}
\eeq
where $p_{\rm hard}(\tau)$ is a time-dependent hard momentum scale and 
$\xi(\tau)$ is a time-dependent parameter reflecting the strength of 
the local momentum-space anisotropy~\cite{Romatschke:2003ms}. We will 
further assume that $f_{\rm iso}$ is a Fermi-Dirac distribution and that
$f_q = f_{{\bar q}}$.

Details of the analytic and numerical calculation of the $\xi$-dependent 
rate will be presented elsewhere. Note that for isotropic 
systems there are appreciable next-to-leading (NLO) order corrections 
for $E/T \lsim 1$ \cite{Thoma:1997dk,Arnold:2002ja,Arleo:2004gn,Turbide:2006mc}. 
When we are sensitive to areas of phase space where there are large 
NLO corrections we will apply $K$-factors to our estimates as indicated.


\section{Space-Time Model}
\label{sec:model}

We now construct a model which interpolates between early-time 
longitudinal free streaming and late-time ideal hydrodynamic 
expansion.  Given (\ref{eq:fansio}) the medium parton energy density 
can be factorized as
\beq
{\cal E}(p_{\rm hard},\xi) \, = \, 
    \int \frac{d^3{\bf p}}{(2\pi)^3} \; p \, f({\bf p},\xi) \, = \, 
    {\cal E}_{0}(p_{\rm hard}) \; {\cal R}(\xi) \;\, ,
\eeq
where ${\cal R}(\xi) = \left[ 1/(\xi+1) + {\rm 
arctan}\sqrt{\xi}/\sqrt{\xi} \right]/2$ and ${\cal 
E}_0(p_{\rm hard})$ is the energy density resulting from integration 
of the isotropic quark and anti-quark distribution 
functions appearing in Eq.~(\ref{eq:fansio}).

{\em Longitudinal Free Streaming Limit}:
For a longitudinally free streaming plasma $\xi_{\rm FS}(\tau) = 
(\tau/\tau_0)^2 -1$ and $p_{\rm hard}$ is constant and equal to the 
initial average hard momentum scale in the plasma; therefore, ${\cal 
E}_{\rm FS}(\tau) = {\cal E}(p_{\rm hard},\xi_{\rm FS}(\tau))$. 
Assuming an isotropic plasma at $\tau=\tau_0$ this results in $p_{\rm 
hard}(\tau) = T_0$ and $\lim_{\tau \gg \tau_0} {\cal E}_{\rm FS} 
\rightarrow {\cal E}_0 \; (\tau_0 / \tau)$ where $T_0$ is the initial 
plasma ``temperature''.  Note that assuming a formation time of 
$\tau_0=0.1$ fm/c at $\tau = 1$ fm/c we have $\xi_{\rm FS} \simeq 
100$.

{\em Ideal Hydrodynamic Expansion}: 
For a plasma which is undergoing ideal longitudinal hydrodynamic 
expansion we have $\xi(\tau)=0$ by assumption. Additionally, since the 
system is thermal we can identify the hard momentum scale with the 
plasma temperature so that $p_{\rm hard}(\tau) = T(\tau) = T_0 
\left(\tau_0 / \tau\right)^{1/3}$. Correspondingly, we have ${\cal 
E}_{\rm hydro} = {\cal E}_0 (\tau_0/\tau)^{4/3}$.

{\em Interpolating Model}:
In order to construct a model which interpolates between longitudinal 
free streaming and hydrodynamic expansion we introduce a smeared step 
function $\lambda(\tau) \equiv \left({\rm 
tanh}\left[\gamma (\tau-\tau_{\rm iso})/\tau_0 \right]+1\right)/2$. 
This allows us to model the time-dependence of $\xi$ and $p_{\rm 
hard}$ as
\bqa
{\cal E}(\tau) &=& {\cal E}_{\rm FS}(\tau) \,
                    \left[\,{\cal U}(\tau)/{\cal U}(\tau_0)\,\right]^{4/3} \; , \nonumber \\
p_{\rm hard}(\tau) &=& T_0 \, \left[\,{\cal U}(\tau)/{\cal U}(\tau_0)\,\right]^{1/3} \; , \nonumber \\
\xi(\tau) &=& a^{2(1-\lambda(\tau))} - 1 \; , 
\label{eq:modelEQs}
\eqa
where ${\cal U}(\tau) \equiv \left[{\cal R}\!\left(a_{\rm iso}^2-
1\right)\right]^{3\lambda(\tau)/4}\left(a_{\rm 
iso}/a\right)^{\lambda(\tau)}$, $a \equiv \tau/\tau_0$ and $a_{\rm 
iso} \equiv \tau_{\rm iso}/\tau_0$.  The power of ${\cal R}$ in ${\cal 
U}$ keeps the energy density continuous at $\tau = \tau_{\rm iso}$ for 
all $\gamma$. 

When $\tau \ll \tau_{\rm iso}$ we have $\lambda \rightarrow 0$ and the 
system is longitudinally free streaming.  When $\tau \gg \tau_{\rm 
iso}$ then $\lambda \rightarrow 1$ and the system is expanding 
hydrodynamically.  In the limit $\gamma \rightarrow \infty$, $\lambda 
\rightarrow \Theta(\tau- \tau_{\rm iso})$.  In 
Fig.~\ref{fig:modelPlot} we plot the time-dependence of ${\cal E}$, 
$p_{\rm hard}$, and $\xi$ assuming $\gamma=2$ for four different 
plasma isotropization times corresponding to $a_{\rm iso} \in 
\{1,4,8,16\}$.


\section{Results}
\label{sec:results}

To obtain the final expected dilepton yields we integrate the 
annihilation rate over $\tau \in \{\tau_0,\tau_f\}$ and $\eta \in \{-
2.5,2.5\}$ with parameters specified by Eq.~(\ref{eq:modelEQs}) and 
$\tau_f$ set by $p_{\rm hard}(\tau_f)=T_c$. In this letter we will 
present expected $e^+e^-$ yields resulting from a Pb-Pb collision at 
LHC full beam energy, $\sqrt{s} = 5.5$ TeV.  At RHIC energies 
sensitivity to $\tau_{\rm iso}$ is reduced due to the poor signal-to-background 
ratio for medium dileptons.  Predictions for Au-Au at RHIC 
energy will be presented elsewhere.

\begin{figure}[t]
\begin{center}
\includegraphics[width=8.5cm]{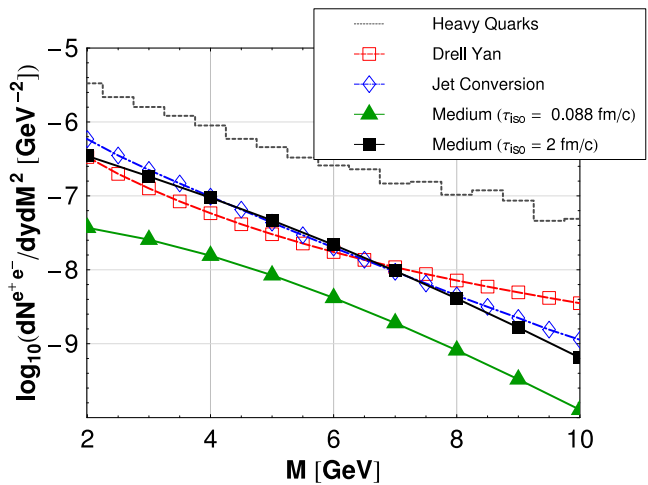}
\end{center}
\vspace{-5.5mm}
\caption{
Dilepton yields as a function of invariant mass with a cut $p_T >$ 8 
GeV.  For medium dileptons we use $\gamma=2$ and $\tau_{\rm iso}$ is 
taken to be either 0.088 fm/c or 2 fm/c. A $K$-factor of 1.5 was 
applied to account for NLO corrections.
}
\label{fig:dileptons-m}
\end{figure}

\begin{figure}[t]
\begin{center}
\includegraphics[width=8.1cm]{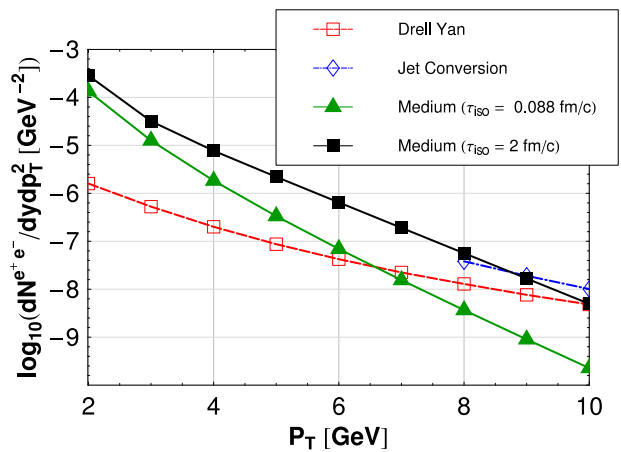}
\end{center}
\vspace{-5.5mm}
\caption{
Dilepton yields as a function of transverse momentum with a cut 0.5 $< 
M <$ 1 GeV.  For medium dileptons we use $\gamma=2$ and $\tau_{\rm 
iso}$ is taken to be either 0.088 fm/c or 2 fm/c. A $K$-factor of 6 
was applied to account for NLO corrections.
}
\label{fig:dileptons-pt}
\end{figure}

\begin{figure}[t]
\begin{center}
\includegraphics[width=8.5cm]{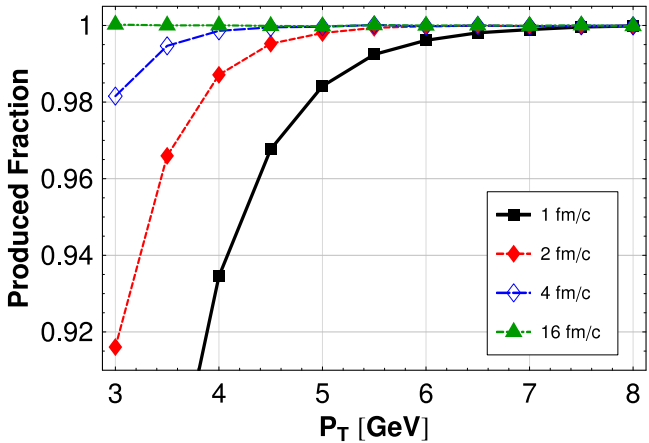}
\end{center}
\vspace{-7mm}
\caption{
Fraction of dileptons produced at $\tau \in$ \{1, 2, 4, 16\} fm/c 
assuming $\tau_{\rm iso}= 0.5$ fm/c. Cuts are the same as 
in Fig.~\ref{fig:dileptons-pt}.  
}
\label{fig:dileptons-pt-time}
\end{figure}

In order to facilitate comparison with previous works we take $\tau_0 
= 0.088$ fm/c, $T_0 = 845$ MeV, $T_c = 160$ MeV, and $R_T = 7.1$ 
fm~\cite{Turbide:2006mc}. Here we assume that when the system reaches 
$T_c$ all medium emission stops. The addition of mixed and hadronic 
phase emission is not included in the present work since 
the kinematic range we consider is not sensitive to the late-time 
evolution of the system (see Fig.~\ref{fig:dileptons-pt-time}).  In 
Figs.~\ref{fig:dileptons-m} and \ref{fig:dileptons-pt} we show our 
final predicted $e^+e^-$ yields as a function of invariant mass and 
transverse momentum along with predicted yields from other sources. 
Predictions for Drell-Yan, heavy quark, and jet conversion yields 
were supplied by the authors of Ref.~\cite{Turbide:2006mc}.

As can be seen from Fig.~\ref{fig:dileptons-m} there is a significant 
variation of the medium dilepton yield when varying the assumed plasma 
isotropization time from $0.088$ fm/c to 2 fm/c.  When an 
isotropization time of 2 fm/c is assumed we see that medium dileptons 
become as important as Drell-Yan and jet conversion.  The reason for 
the enhanced production is that longitudinal free streaming preserves 
more transverse momentum than an hydrodynamically expanding plasma. 
However, all three contributions are down by an order of magnitude 
from the expected background coming from semileptonic heavy quark 
decay.  In practice this would mean that experimentalists would have 
to subtract this background to 10\%.  We note that an indefinitely 
longitudinally free streaming plasma ($\tau_{\rm iso}\rightarrow 
\infty$) produces less low energy ($M, p_T \lsim 2$ GeV) dileptons due 
to the rapidly dropping parton densities as $\xi\rightarrow\infty$.

As we show in Fig.~\ref{fig:dileptons-pt} as a function of $p_T$ the 
medium contribution dominates the expected Drell-Yan and jet 
conversion sources for all $p_T \lsim 6$ GeV.  If an isotropization 
time of 2 fm/c is assumed then the medium dileptons dominate out to 
$p_T \sim 9$ GeV.  This dominance means that it should be possible to 
use dilepton production to determine much-needed information about 
quark-gluon plasma initial conditions at LHC.  As can be seen from 
Fig.~\ref{fig:dileptons-pt} at $p_T =$ 5 GeV the expected medium 
dilepton yield varies by nearly an order of magnitude depending on the 
assumed plasma isotropization time.  This level of variation will 
hopefully be measurable at LHC.

\begin{figure}[t]
\begin{center}
\includegraphics[width=8.5cm]{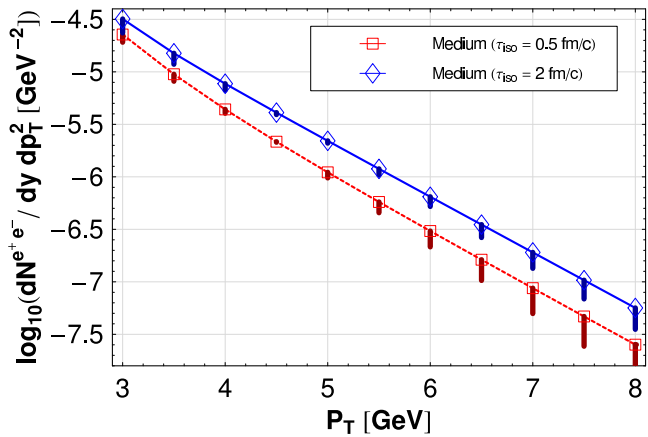}
\end{center}
\vspace{-7mm}
\caption{
Dilepton yields as a function of transverse momentum.  Shown are 
yields obtained assuming $\tau_{\rm iso} = 0.5$ fm/c and 2 fm/c  
with error bars indicating model variation, $0.05 \leq \gamma \leq 10$.  
Cuts and $K$-factor are the same as in Fig.~\ref{fig:dileptons-pt}.
}
\label{fig:dileptons-pt-b}
\end{figure}

To illustrate the dependence on the model parameter $\gamma$ in 
Fig.~\ref{fig:dileptons-pt-b} we have plotted medium dilepton yields 
obtained assuming $\tau_{\rm iso}=0.5$ fm/c and $\tau_{\rm iso}=2$ 
fm/c.  The central values obtained are with $\gamma=2$ and the error bars 
come from variation of $\gamma$ in the range $0.05 < \gamma < 10$.  As 
can be seen from this Figure between 3 and 8 GeV there is little 
sensitivity to the parameter $\gamma$.

In Fig.~\ref{fig:dileptons-pt-time} we show snapshots of the fraction 
of medium dileptons produced as a function of transverse momentum. 
What this figure shows is that by 4 fm/c yields in this kinematic 
regime are saturated. At 1 fm/c approximately 94\% of all $p_T=4$ GeV 
dileptons have already been produced as well as 98\% of the $p_T=5$ GeV 
pairs.  This highlights the sensitivity of this observable to early-times 
after a heavy-ion collision and justifies neglecting the effect 
of transverse expansion and mixed/hadronic phases when considering 
dileptons in this kinematic regime. 

\section{Conclusions and Discussion}
\label{sec:conclusions}

Based on Figs.~\ref{fig:dileptons-pt} and \ref{fig:dileptons-pt-b} it 
should be possible to measure $\tau_{\rm iso}$ at LHC energies 
using dilepton production in the kinematic range $3 < p_T < 8$ GeV. 
Additionally via our model for $\xi(\tau)$ given in 
Eq.~\ref{eq:modelEQs} determining $\tau_{\rm iso}$ provides an 
estimate of the maximum amount of momentum-space anisotropy achieved 
during the lifetime of the QGP. 

The effect of varying $\tau_{\rm iso}$ is also large in the dilepton 
spectra vs invariant mass as shown in Fig.~\ref{fig:dileptons-m}, 
however, Drell-Yan and jet conversion production can be up to 10 times 
larger than medium production making it difficult to measure a clean 
medium dilepton signal.

Our chief uncertainty is the NLO order corrections to dilepton 
production incorporating anisotropies. These corrections are 
particularly important for low-mass dilepton production.  At LHC 
energy it is possible to reduce sensitivity to these NLO corrections 
by placing cuts $M,p_T \gsim 2$ GeV.  Another uncertainty comes from 
our assumption of chemical equilibrium. Naively finite chemical 
potentials should affect isotropic and anisotropic plasmas equally so 
one expects that although the total yields could change one would 
still see a sensitivity to the assumed isotropization/thermalization 
time.  At leading order in the quark fugacity, $\lambda_q$, the ratio of the 
isotropic to anisotropic rates should be independent 
of $\lambda_{q}$ \cite{Strickland:1994rf}.

Future work will study the effect of finite quark chemical potentials,
collisional broadening of the parton distributions, and the 
possibility of late-time persistent anisotropies (finite viscosity).
In addition, models such as (\ref{eq:modelEQs}) can be used to assess
the impact of momentum-space anisotropies on other observables.


\vspace{-2mm}
\section*{Acknowledgments}
We thank A.~Dumitru, A.~Ipp, B.~Schenke, and S.~Turbide.
M.M. was supported by the Helmholtz Research School and
M.S. by DFG project GR 1536/6-1.


\vspace{-1mm}
\bibliography{anisodileptons}


\end{document}